\documentclass[a4paper,10pt,twoside]{cpc-hepnp}
\usepackage{CJK,upgreek,fancyhdr}
\usepackage{multicol}
\usepackage{graphicx}
\usepackage{booktabs}
\usepackage{amssymb,bm,mathrsfs,bbm,amscd}
\usepackage[tbtags]{amsmath}
\usepackage{lastpage}

\begin{document}
\begin{CJK*}{GB}{gbsn}

\fancyhead[c]{\small Chinese Physics C~~~Vol. xx, No. x (2020) xxxxxx}
\fancyfoot[C]{\small 010201-\thepage}

\footnotetext[0]{Received 7 Feb 2020}

\title{Accurate correction of arbitrary spin fermions\\
quantum tunneling from non-stationary Kerr-de Sitter black hole\\
based on corrected Lorentz dispersion relation
\thanks{This work is supported by National Natural Science
Foundation of China (11573022), National Natural Science Foundation of China (11273020), and Natural Science Foundation of Shandong Province, China (ZR2019MA059). }}

\author{%
      Bei Sha(ɳ±´)$^{1;1)}$\email{shabei@qlnu.edu.cn}%
\quad Zhi-E Liu(ÁõÖ¾¶ð)$^{1;2)}$\email{zhieliu@163.com}%
\quad Yu-Zhen Liu(ÁõÓñÕæ)$^{1;3)}$\\\email{liuyz249@sina.com}%
\quad Xia Tan(̷ϼ)$^{1;4)}$\email{adfyt@163.com}
\quad Jie Zhang(ÕŽà)$^{1;5)}$\email{327093017@qq.com}%
\quad Shu-Zheng Yang(ÑîÊ÷Õþ)$^{2;6)}$\email{szyangphys@126.com}%
}
\maketitle

\address{%
$^1$ School of Physics and Electronic Engineering, Qilu Normal University, Jinan 250200, China
}
\address{%
$^2$ College of Physics and Space Science, China West Normal University, Nanchong 637002, China
}

\begin{abstract}
According to a corrected dispersion relation proposed in the study on string theory and quantum gravity theory, Rarita-Schwinger equation has been precisely modified, which results in a Rarita-Schwinger-Hamilton-Jacobi equation, and through which, the characteristics of arbitrary spin fermions quantum tunneling radiation from non-stationary Kerr-de Sitter black hole are researched. A series of accurately corrected physical quantities such as surface gravity, chemical potential, tunneling probability and Hawking temperature that describe the properties of the black hole are derived. This research has enriched the research methods and made precision of the research contents of black hole physics.
\end{abstract}

\begin{keyword}
Black hole,   Fermions,   Tunneling radiation,   Lorentz dispersion relation
\end{keyword}

\begin{pacs}
 04.70.DY, 04.62.+V, 11.30.-j
\end{pacs}


\begin{multicols}{2}

\section{Introduction}

Hawking found that black holes radiated particles outward its event horizon through introduction of quantum theory. This kind of radiation was called Hawking radiation. Hawking radiation theory profoundly reveals the inner link among the gravitation theory, the quantum theory and the thermodynamic statistical physics. Hawking did not use the concepts of quantum tunneling and potential barrier in the specific calculation, nor did he consider the influences of the particles emission to the black hole's mass, event horizon as such the black hole characteristic quantities, though he had used the concept of quantum tunneling effect when proving black hole radiation~\citep{lab1Hawking1974BlackN,lab2Hawking1975ParticleCMP}. Such a calculation resulted in that Hawking radiation had a precise form of Planck blackbody radiation, also led to the difficulty of black hole ``information loss". Such an information loss (non-conservation) would lead to a major crisis in quantum theory.

Hawking¡¯s theory of tunneling radiation raised people's enthusiasm for study on black holes. Kraus, Parikh, Wilczek et al. studied the quantum tunneling radiation of black holes by considering the energy conservation conditions in the radiation process, and explained Hawking's thermal radiation well. This theory not only revised Hawking's pure thermal radiation theory, but also promoted the research and development of black hole physics~\citep{lab3TM-Kraus1995Self-interactionNPB,lab4TM-Parikh2000HawkingPRL,lab5TM-Zhao2006HowCPL,lab6TM-Ren2006TunnellingCPL,
lab7TM-Liu2009ChargedPLB,lab8TM-Fang2005ParticleIJMPD,lab9TM-Lin2008HawkingCPL,lab10TM-Banerjee2009ConnectingPRD}. Then semi-classical Hamilton-Jacobi method was proposed to research black holes¡¯ tunneling radiation~\citep{lab11HJ-Srinivasan1999ParticlePRD,lab12HJ-Shankaranarayanan2002HawkingCQG}. Through this method Klein-Gordon equation descripting the behavior of the particles with spin 0 was rewritten as the form of action, and then was expanded using WKB approximation, ignored all the terms with $\hbar$ which was considered as small quantity, the semi-classical Hamilton-Jacobi equation (Hereinafter refers to as H-J equation)  was gotten. In 2007, Kenner and Mann et al. firstly used the semi-classical theory to study the quantum tunneling of fermions with spin 1/2~\citep{lab13HJ-Kenner2008FermionsCQG,lab14HJ-Kenner2008ChargedPLB}. They decomposed the field equation into two cases of spin up and spin down, and obtained the tunneling rate of the Dirac particles and the temperature of black holes. Yang and Lin developed their theory and method. They derived the H-J equation from Dirac equation and Rarita-Schwinge equation by selecting the Gamma matrix and applying the semi-classical theory in higher-dimensional and lower-dimensional curved space-time. That is to say, the H-J equation can uniformly describe the dynamic behavior of arbitrary spin fermions in curved space-time, and the calculation process was simplified, thus the calculation amount was greatly reduced by H-J method~\citep{lab15HJ-Lin2009FermionPRD,lab16HJ-Lin2009FermionsPLB}.

Recent research on the quantum gravity theory suggested that the Lorentz dispersion relation needed to be modified in high energy case~\citep{lab17LV-Amelino-Camelia2002RelativityIJMPD,lab18LV-Magueijo2002LorentzPRL,lab19LV-Ellis1992StringPLB,
lab20LV-Ellis2004SynchrotronAPP,lab21LV-Jacobson2003strongN,lab22LV-Kruglov2013ModifiedMPLA}. Accordingly, the quantum tunneling radiation at the event horizon of the black hole has to be modified due to the conservation of energy and momentum. Since the H-J equation in the curved space-time corresponds to the Lorentz relation of energy and momentum in the curved space-time, the correction of the Lorentz dispersion relation will inevitably correct the H-J equation of the particles in the strong gravitational field, which will lead to the correction of physical quantities such as fermions or bosons quantum tunneling rate and black hole temperature. Although the modification of the Lorentz dispersion relation is only a small correction on the Planck scale, it has a significant impact on the quantum tunneling radiation of the black hole~\citep{lab23ELV-Bailey2015Short-rangePRD,lab24ELV-Kostelecky1991CPT-NLB,lab25ELV-Kostelecky1995CPT-PRD,
lab26ELV-Kostelecky1989SpontaneousPRD,lab27ELV-Kostelecky2009ProspectsPRL}. Accurately correcting the tunneling of arbitrary spin fermions at the event horizon of a black hole is a subject that needs further study.

Some studies have been done on the correction of the tunneling radiation of static and stationary black holes~\citep{lab28SBH-Damour1976Black-holePRD,lab29SBH-Sannan1988HeuristicGRG,
lab30SBH-Robinson1975UniquenessPRL,lab31SBH-Gibbons1978BlackPRSA,lab32SBH-Zhao1983HawkingCAA,
lab33SBH-Unru1976NotesPRD,lab34SBH-Hartle1976Path-integralPRD,lab35SBH-Sha2020 AccurateAHEP,
lab36SBH-Sha2020HawkingAA,lab37SBH-Liu2019CorrectionMPLA}. Actually real black holes in the universe show as non-stationary black holes due to the action of evaporation, accretion or mergers. For non-stationary black holes, Zhao et al. found a method to determine the black hole temperature at the event horizon -- tortoise coordinate transformation~\citep{lab38NBH-Zhao1991HawkingCPL}. Subsequently, a series of studies were conducted on non-stationary black holes
~\citep{lab40NBH-Zhao1992HawkingCPL,lab39NBH-Zhao1994Damour-RuffiniIJTP,lab41NBH-Li1994HawkingCPL,lab42NBH-Sun1995HawkingILNCB,lab43NBH-Yang1995dependenceAPS}. However, few studies have been reported on the correction of quantum tunneling radiation from non-stationary black holes based on the corrected Lorentz dispersion relation, which is a kind of subject worthy of studying. In this paper, the arbitrary spin fermions quantum tunneling from non-stationary Kerr-de Sitter black hole will be researched based on the corrected Lorentz dispersion relation.

The contents of this paper are as follows. In section 2, the dynamics equation of fermions in the space-time of non-stationary Kerr-de Sitter black hole will be derived based on the corrected Lorentz dispersion relation. Some precisely corrected, important physical quantities describing the quantum tunneling characteristics of arbitrary spin fermions, subsequently the black hole temperature will be obtained in section 3. In section 4, the conclusions and the related discussions of this study including black hole entropy will be given.
\\

\section{Correction of the dynamics equation for arbitrary spin fermions in non-stationary Kerr-de Sitter black hole space-time}

The modified Lorentz dispersion relation proposed in string theory and quantum gravity theory is ~\citep{lab44LV-Magueijo2003Generalized,lab45LV-Amelino-Camelia2004PhenomenologyNJP,lab18LV-Magueijo2002LorentzPRL,
lab19LV-Ellis1992StringPLB,lab20LV-Ellis2004SynchrotronAPP,lab21LV-Jacobson2003strongN,
lab22LV-Kruglov2013ModifiedMPLA}

\begin{eqnarray}
\label{eq1}
p_{0}^{2}=p^{2}+m^{2}-(Lp_{0})^{\alpha}p^{2},
\end{eqnarray}
where $L$ is a constant on the Planck scale. When   is considered, the general fermions¡¯ dynamic equation R-S equation~\citep{lab46EqRS-Rarita1941 theoryPR,lab47EqRS-Gibbons1976noteJPA} can be extended to Kerr-de Sitter curved space-time as~\citep{lab48EqRS-Yang2019ModifiedSSPMA,lab35SBH-Sha2020 AccurateAHEP,
lab37SBH-Liu2019CorrectionMPLA}
\begin{eqnarray}
\label{eq2}
(\gamma^{\mu}D_{\mu}+\frac{m}{\hbar}-\sigma\hbar\gamma^{t}D_{t}\gamma^{j}D_{j})\psi_{\alpha_{1}\cdots\alpha_{k}}=0,
\end{eqnarray}
where $\gamma^{\mu}$ is the Gamma matrix in curved space-time and satisfies the following condition
\begin{eqnarray}
\label{eq2-1}
\{\gamma^{\mu},\gamma^{\nu}\}=2g^{\mu\nu}I,
\end{eqnarray}
and $D_{\mu}$ is the covariant derivative operation symbol of curved space-time, that is
\begin{eqnarray}
\label{eq3}
D_{\mu}=\partial_{\mu}+\Omega_{\mu}+\frac{i}{\hbar}eA_{\mu},
\end{eqnarray}
where $\Omega_{\mu}$ is the rotational connection in curved space-time. As quantum scale corrections, $0<\alpha\ll1$ , so $\sigma\hbar\gamma^{t}D_{t}\gamma^{j}D_{j}\psi_{\alpha_{1}\cdots\alpha_{k}}$ is a small term. This matrix equation can only be solved in the specific curved space-time, so the fermions¡¯ wave function is set first as
\begin{eqnarray}
\label{eq4}
\psi_{\alpha_{1}\cdots\alpha_{k}}=\xi_{\alpha_{1}\cdots\alpha_{k}}e^{\frac{i}{\hbar}S},
\end{eqnarray}
where $S$ is the action of fermions with mass $m$ in the space-time of Kerr-de Sitter black hole.

In order to solve it, the ~Eq.~(\ref{eq2}) is rewritten as
\begin{eqnarray}
\label{eq5}
(i\gamma^{\mu}\partial_{\mu}S+m+\sigma\gamma^{\nu}\partial_{\nu}\gamma^{j}\partial_{j}S)\xi_{\alpha_{1}\cdots\alpha_{k}}=0,
\end{eqnarray}
where $\mu=0,1,2,3; j=1,2,3$. For a non-stationary black hole, we use the advanced Eddington coordinate $v$ to represent the dynamic characteristics. To solve ~Eq.~(\ref{eq5}), defining
\begin{eqnarray}
\label{eq6}
\nonumber\Gamma^{\mu}&=&i\gamma^{\mu}+\sigma\partial_{v}S\gamma^{v}\gamma^{\mu}\\[1mm]
\Gamma^{\nu}&=&i\gamma^{\nu}+\sigma\partial_{v}S\gamma^{v}\gamma^{\nu}\\[1mm]
\nonumber m_{D}&=&m-\sigma g^{vv}(\partial_{v}S)^{2},
\end{eqnarray}
so ~Eq.~(\ref{eq5}) becomes
\begin{eqnarray}
\label{eq7}
(\Gamma^{\mu}\partial_{\mu}S+m_{D})\xi_{\alpha_{1}\cdots\alpha_{k}}=0.
\end{eqnarray}
Multiplying both sides of~Eq.~(\ref{eq7}) by $\Gamma^{\mu}\partial_{\mu}S$, and we will get,
\begin{eqnarray}
\label{eq8}
(\Gamma^{\nu}\Gamma^{\mu}\partial_{\nu}S\partial_{\mu}S-m_{D}^{2})\xi_{\alpha_{1}\cdots\alpha_{k}}=0,\\[2mm]
\label{eq9}
(\Gamma^{\mu}\Gamma^{\nu}\partial_{\mu}S\partial_{\nu}S-m_{D}^{2})\xi_{\alpha_{1}\cdots\alpha_{k}}=0.
\end{eqnarray}
Eq.~(\ref{eq8}) and~(\ref{eq9}) are equivalent. Adding~Eq.~(\ref{eq8}) and~(\ref{eq9}) and considering~Eq.~(\ref{eq6})and~(\ref{eq2-1}), we can get
\end{multicols}
\ruleup
\begin{equation}
\label{eq10}
[g^{\mu\nu}\partial_{\mu}S\partial_{\nu}S-2i\sigma\partial_{v}Sg^{v\nu}\partial_{\nu}S\gamma^{\mu}\partial_{\mu}S
-\sigma^{2}(\partial_{v}Sg^{v\nu}\partial_{\nu}S)^{2}+m^{2}-2m\sigma g^{vv}(\partial_{v}S)^{2}
+\sigma^{2} (g^{vv})^{2}(\partial_{v}S)^{4}]\xi_{\alpha_{1}\cdots\alpha_{k}}=0.
\end{equation}
Dividing both sides of~Eq.~(\ref{eq10}) by $-2\partial_{v}Sg^{v\nu}\partial_{\nu}S$, and we get
\begin{equation}
\label{eq11}
\left\{i\sigma\gamma^{\mu}\partial_{\mu}S-\frac{g^{\mu\nu}\partial_{\mu}S\partial_{\nu}S-\sigma^{2}(\partial_{v}Sg^{v\nu}\partial_{\nu}S)^{2}+m^{2}-2m\sigma g^{vv}(\partial_{v}S)^{2}+\sigma^{2}(g^{vv})^{2}(\partial_{v}S)^{4}}{2\partial_{v}S g^{v\nu}\partial_{\nu}S}\right\}\xi_{\alpha_{1}\cdots\alpha_{k}}=0.
\end{equation}
Defining
\begin{equation}
\label{eq12}
m_{l}=-\frac{g^{\mu\nu}\partial_{\mu}S\partial_{\nu}S-\sigma^{2}(\partial_{v}Sg^{v\nu}\partial_{\nu}S)^{2}+m^{2}-2m\sigma g^{vv}(\partial_{v}S)^{2}
+\sigma^{2}(g^{vv})^{2}(\partial_{v}S)^{4}}{2\partial_{v}Sg^{v\nu}\partial_{\nu}S},
\end{equation}
then~Eq.~(\ref{eq10}) becomes

\begin{multicols}{2}

\begin{equation}
\label{eq13}
(i\sigma\gamma^{\mu}\partial_{\mu}S+m_{l})\xi_{\alpha_{1}\cdots\alpha_{k}}=0.
\end{equation}
Multiplying both sides of~Eq.~(\ref{eq13}) by $i\sigma\gamma^{\nu}\partial_{\nu}S$, and we get
\begin{equation}
\label{eq14}
(\sigma^{2}\gamma^{\mu}\gamma^{\nu}\partial_{\mu}S\partial_{\nu}S+m_{l}^{2})\xi_{\alpha_{1}\cdots\alpha_{k}}=0.
\end{equation}
In~Eq.~(\ref{eq14}), $\mu$ and $\nu$ are interchanged and~Eq.~(\ref{eq15}) can be obtained
\begin{equation}
\label{eq15}
(\sigma^{2}\gamma^{\nu}\gamma^{\mu}\partial_{\nu}S\partial_{\mu}S+m_{l}^{2})\xi_{\alpha_{1}\cdots\alpha_{k}}=0.
\end{equation}
Adding~Eq.~(\ref{eq14}) and~Eq.~(\ref{eq15}) and combining with~Eq.~(\ref{eq2-1}) , we obtain
\begin{equation}
\label{eq16}
(\sigma^{2}g^{\mu\nu}\partial_{\mu}S\partial_{\nu}S+m_{l}^{2})\xi_{\alpha_{1}\cdots\alpha_{k}}=0.
\end{equation}
Eq.~(\ref{eq16}) is a matrix equation. In fact, it is an eigenvalue equation, the condition of which has a non-zero solution is that the value of its coefficient determinant is equal to zero. That is
\end{multicols}
\ruleup
\begin{equation}
\label{eq17}
\sigma^{2}g^{\mu\nu}\partial_{\mu}S\partial_{\nu}S+\left\{-\dfrac{g^{\mu\nu}\partial_{\mu}S\partial_{\nu}S-\sigma^{2}(\partial_{v}Sg^{v\nu}\partial_{\nu}S)^{2}+m^{2}-2m\sigma g^{vv}(\partial_{v}S)^{2}
+\sigma^{2}(g^{vv})^{2}(\partial_{v}S)^{4}}{2\partial_{v}Sg^{v\nu}\partial_{\nu}S}\right\}^{2}=0.
\end{equation}
Taking notice of $g^{\mu\nu}\partial_{\mu}S\partial_{\nu}S=-m^{2}$, the~Eq.~(\ref{eq17}) can become
\begin{equation}
\label{eq18}
g^{\mu\nu}\partial_{\mu}S\partial_{\nu}S+m^{2}-2m\sigma g^{vv}(\partial_{v}S)^{2}-2m\sigma\partial_{v}Sg^{v\nu}\partial_{\nu}S-\sigma^{2}(\partial_{v}Sg^{v\nu}\partial_{\nu}S)^{2}
+\sigma^{2}(g^{vv})^{2}(\partial_{v}S)^{4}=0.
\end{equation}

\begin{multicols}{2}
\vspace{0.5cm}
We have maintained the correction item of $\sigma$ in~Eq.~(\ref{eq18}), so~Eq.~(\ref{eq18}) is the precise correction of R-S equation considering the correction of Lorentz dispersion relation. Obviously, it is actually also an accurate correction of H-J equation, we call~Eq.~(\ref{eq18}) as Rarita-Schwinger-Hamilton-Jacobi equation (R-S-H-J equation for short). Following, we will use this precisely corrected R-S-H-J equation to study the dynamic behavior of arbitrary spin fermions in non-stationary Kerr-de Sitter black hole space-time and thereby the properties of the black hole.

\section{Accurate correction of arbitrary spin fermions tunneling in the space-time of non-stationary Kerr-de Sitter black hole}

In the advanced Eddington-Finkelstein coordinates, the line element of the non-stationary Kerr-de Sitter black hole can be written as~\citep{lab49LE-Carmeli1977GravitationalAP,lab50LE-Xu1998RadiatingCQG}
\begin{eqnarray}
\label{eq19}
&&ds^{2}\nonumber\\
&&
=A\nonumber\frac{1}{\Sigma}(\Delta_{\lambda}-\Delta_{\theta}a^{2}\sin^{2}\theta)dv^{2}\\[1mm]
&&
-2\sqrt{A}(dv-a\sin^{2}\theta d\varphi)dr-\frac{\Sigma}{\Delta_{\theta}}d\theta^{2}\\[1mm]
&&
+\nonumber A\frac{2a}{\Sigma}[\Delta_{\theta}(r^{2}+a^{2})-\Delta_{\lambda}]\sin^{2}\theta dvd\varphi\\[1mm]
&&
-\nonumber A\frac{1}{\Sigma}[\Delta_{\theta}(r^{2}+a^{2})-\Delta_{\lambda}\sin^{2}\theta]\sin^{2}\theta d\varphi^{2},
\end{eqnarray}
where
\begin{eqnarray}
\label{eq20}
A\nonumber&=&(1+\frac{1}{3}\Lambda a^{2})^{-2}\\
\nonumber\Sigma&=&r^{2}+a^{2}\cos^{2}\theta\\
\Delta_{\lambda}&=&r^{2}+a^{2}-2Mr-\frac{1}{3}\Lambda r^{2}(r^{2}+a^{2})\\
\nonumber\Delta_{\theta}&=&1+\frac{1}{3}\Lambda a^{2}\cos^{2}\theta\\
M&=&M(v)\nonumber\\
a&=&a(v)\nonumber.
\end{eqnarray}
$M$ is the mass of the black hole, and   is the angular momentum per unit mass of the black hole. From formula (\ref{eq19}), we can find the non-zero components of the contravariant metric tensor of Kerr-de Sitter black hole are as follows
\begin{eqnarray}
\label{eq21}
g^{00}\nonumber&=&-\frac{a^{2}\sin^{2}\theta}{\Delta_{\theta}A\Sigma}\\
g^{11}\nonumber&=&-\frac{\Delta_{\lambda}}{\Sigma}\\
g^{22}\nonumber&=&-\frac{\Delta_{\theta}}{\Sigma}\\
g^{33}&=&-\frac{1}{\Delta_{\theta}A\Sigma\sin^{2}\theta}\\
g^{01}\nonumber&=&g^{10}=-\frac{r^{2}+a^{2}}{\sqrt{A}\Sigma}\\
g^{03}\nonumber&=&g^{30}=-\frac{a}{\Delta_{\theta}A\Sigma}\\
g^{13}\nonumber&=&g^{31}=-\frac{a}{\sqrt{A}\Sigma},
\end{eqnarray}
According to the zero hypersurface condition, the event horizon equation satisfies
\begin{eqnarray}
\label{eq22}
g^{\mu\nu}\frac{\partial f}{\partial x_{\mu}}\frac{\partial f}{\partial x_{\nu}}=0,
\end{eqnarray}
where $f$ is a hypersurface. Since the space-time of Kerr-de Sitter black hole is axisymmetric,~Eq.~(\ref{eq22}) is independent of $\varphi$, so $\frac{\partial f}{\partial\varphi}=0$ , and the equation of event horizon of Kerr-de Sitter black hole can be expressed as
\begin{eqnarray}
\label{eq23}
\nonumber&&g^{00}\left\{\frac{\partial f}{\partial v}\right\}^{2}+g^{11}\left\{\frac{\partial f}{\partial r}\right\}^{2}\\
&&+g^{22}\left\{\frac{\partial f}{\partial \theta}\right\}^{2}+2g^{01}\frac{\partial f}{\partial v}\frac{\partial f}{\partial r}=0.
\end{eqnarray}
Its zero hypersurface is
\begin{eqnarray}
\label{eq24}
f=f(r,v,\theta)=0,
\end{eqnarray}
where $r$ is the function of $v$ and $\theta$ , $r=r(v,\theta)$ . In order to obtain the position of event horizon of the black hole, we need calculate the rate of change of $f$ with respect to each of its components. Taking the partial derivative of~Eq.~(\ref{eq24}), we get
\begin{eqnarray}
\label{eq25}
\nonumber&&\frac{\partial f}{\partial v}=-\frac{\partial f}{\partial r}\frac{\partial r}{\partial v}\\[1mm]
&&\frac{\partial f}{\partial \theta}=-\frac{\partial f}{\partial r}\frac{\partial r}{\partial \theta}.
\end{eqnarray}
So the~Eq.~(\ref{eq23}) becomes
\begin{eqnarray}
\label{eq26}
g^{00}\left\{\frac{\partial r}{\partial v}\right\}^{2}+g^{11}+g^{22}\left\{\frac{\partial r}{\partial \theta}\right\}^{2}+2g^{01}\frac{\partial r}{\partial v}=0.
\end{eqnarray}
Substituting $g^{00},g^{11},g^{22},g^{01}$  into~Eq.~(\ref{eq26}), when $r\rightarrow r_{H}$ , the equation of event horizon is obtained as follows
\begin{eqnarray}
\label{eq27}
\nonumber&&a^{2}\sin^{2}\theta \dot{r_{H}}^{2}-2\Delta_{\theta}\sqrt{A}(r^{2}+a^{2})\dot{r_{H}}\\[1mm]
&&+\Delta_{\lambda}\Delta_{\theta}A+\Delta_{\theta}^{2}A r_{H}^{'2}=0,
\end{eqnarray}
where
\begin{eqnarray}
\label{eq28}
\nonumber&&\dot{r}_{H}=\frac{\partial r_{H}}{\partial v}\\[1mm]
&&r_{H}^{'}=\frac{\partial r_{H}}{\partial \theta},
\end{eqnarray}
which represents the change of the position of the event horizon with time and angle. For Kerr-de Sitter black hole, substituting~Eq.~(\ref{eq21})into~Eq.~(\ref{eq18}) -- the accurately corrected R-S-H-J equation, the specific expression of~Eq.~(\ref{eq18}) is
\begin{eqnarray}
\label{eq29}
\nonumber&&g^{00}\left\{\frac{\partial S}{\partial v}\right\}^{2}+g^{11}\left\{\frac{\partial S}{\partial r}\right\}^{2}+g^{22}\left\{\frac{\partial S}{\partial \theta}\right\}^{2}+g^{33}\left\{\frac{\partial S}{\partial \varphi}\right\}^{2}\\[1mm]
\nonumber&&+2g^{01}\frac{\partial S}{\partial v}\frac{\partial S}{\partial r}+2g^{03}\frac{\partial S}{\partial v}\frac{\partial S}{\partial \varphi}+2g^{13}\frac{\partial S}{\partial r}\frac{\partial S}{\partial \varphi}\\[1mm]
&&+m^{2}-2m\sigma\frac{\partial S}{\partial v}\left\{g^{00}\frac{\partial S}{\partial v}+g^{01}\frac{\partial S}{\partial r}+g^{03}\frac{\partial S}{\partial \varphi}\right\}\\[1mm]
\nonumber&&-2m\sigma g^{00}\left\{\frac{\partial S}{\partial v}\right\}^{2}+\sigma^{2}(g^{00})^{2}\left\{\frac{\partial S}{\partial v}\right\}^{4}\\[1mm]
\nonumber&&+\sigma^{2}\left\{\frac{\partial S}{\partial v}\right\}^{2}\left\{\left\{g^{00}\frac{\partial S}{\partial v}\right\}^{2}+\left\{g^{01}\frac{\partial S}{\partial r}\right\}^{2}+\left\{g^{03}\frac{\partial S}{\partial \varphi}\right\}^{2}\right\}\\[1mm]
\nonumber&&=0.
\end{eqnarray}
Since the event horizon of the black hole varies with time, it requires doing a generalized tortoise coordinate transformation to solve the tunneling probability of Fermions from the event horizon of the black hole. The Kerr-de Sitter black hole is an axisymmetric black hole, so we take the following transformation
\begin{eqnarray}
\label{eq30}
\nonumber&&r_{\ast}=r+\frac{1}{2\kappa}\ln \frac{r-r_{H}(v,\theta)}{r_{H}(v_{0},\theta_{0})}\\[1mm]
&&v_{\ast}=v-v_{0}\\[1mm]
\nonumber&&\theta_{\ast}=\theta-\theta_{0}.
\end{eqnarray}
Taking the partial derivative of~Eq.~(\ref{eq30}), we get
\begin{eqnarray}
\label{eq31}
\nonumber&&\frac{\partial S}{\partial r}=\frac{\partial S}{\partial r_{\ast}}\frac{2\kappa[r-r_{H}(v,\theta)]+1}{2\kappa[r-r_{H}(v,\theta)]}\\[1mm]
&&\frac{\partial S}{\partial v}=\frac{\partial S}{\partial v_{\ast}}-\frac{\partial S}{\partial r_{\ast}}\frac{\dot{r}_{H}}{2\kappa[r-r_{H}(v,\theta)]}\\[1mm]
\nonumber&&\frac{\partial S}{\partial \theta}=\frac{\partial S}{\partial \theta_{\ast}}-\frac{\partial S}{\partial r_{\ast}}\frac{r_{H}^{'}}{2\kappa[r-r_{H}(v,\theta)]},
\end{eqnarray}
where
\begin{eqnarray}
\label{eq32}
&&\dot{r}_{H}(v,\theta)=\frac{\partial r_{H}(v,\theta)}{\partial v},\\[1mm]
&&\nonumber r_{H}^{'}(v,\theta)=\frac{\partial r_{H}(v,\theta)}{\partial\theta},
\end{eqnarray}
which represent that the position of the event horizon varies with time and angle. In the Kerr-de Sitter black hole space-time, the action of arbitrary spin fermions can be expressed as follows
\begin{eqnarray}
\label{eq33}
S=S(v_{\ast},r_{\ast},\theta_{\ast},\varphi).
\end{eqnarray}
Though $S$ can't be separated, it is for sure that
\begin{eqnarray}
\label{eq34-1}
&&\frac{\partial S}{\partial v_{\ast}}=-\omega,\\[1mm]
\nonumber &&\frac{\partial S}{\partial \varphi}=n,
\end{eqnarray}
and let
\begin{eqnarray}
\label{eq34-2}
&&\frac{\partial S}{\partial \theta_{\ast}}=P_{\theta_{\ast}}.
\end{eqnarray}
Eq.~(\ref{eq29}) is further sorted out and simplified by writing the second-order small quantity as $O(\sigma^{2})$ , which is not let to participate in the separation of variables, and we get
\begin{eqnarray}
\label{eq35}
\nonumber&&(1-4m\sigma)g^{00}\left\{\frac{\partial S}{\partial v}\right\}^{2}+g^{11}\left\{\frac{\partial S}{\partial r}\right\}^{2}\\[1mm]
&&+g^{22}\left\{\frac{\partial S}{\partial \theta}\right\}^{2}+g^{33}\left\{\frac{\partial S}{\partial \varphi}\right\}^{2}\\[1mm]
&&\nonumber+2(1-2m\sigma)g^{01}\frac{\partial S}{\partial v}\frac{\partial S}{\partial r}+2(1-m\sigma)g^{03}\frac{\partial S}{\partial v}\frac{\partial S}{\partial \varphi}\\[1mm]
&&\nonumber+2g^{13}\frac{\partial S}{\partial r}\frac{\partial S}{\partial \varphi}+m^{2}+O(\sigma^{2})=0.
\end{eqnarray}
Substituting formulas~(\ref{eq31}),~(\ref{eq32}),~(\ref{eq34-1}) and~(\ref{eq34-2}) into~Eq.~(\ref{eq35}), we get (In order not to make the formula too long, the symbol (v,¦È) of the independent variables in the following  representation  will be omitted.)
\end{multicols}
\ruleup
\begin{eqnarray}
\label{eq36}
&&\nonumber\frac{(1-4m\sigma)g^{00}\dot{r_{H}}^{2}+g^{11}\{2\kappa[r-r_{H}]+1\}^{2}+g^{22}r_{H}^{'2}
-2(1-m\sigma)g^{01}\{2\kappa[r-r_{H}]+1\}\dot{r}_{H}}{2\kappa[r-r_{H}]}\left\{\frac{\partial S}{\partial r_{\ast}}\right\}^{2}\\[1mm]
&&\nonumber-\{-2(1-4m\sigma)g^{00}\dot{r_{H}}+2(1-m\sigma)g^{01}\{2\kappa[r-r_{H}]+1\}\}\omega\frac{\partial S}{\partial r_{\ast}}\\[1mm]
&&+\{-2(1-m\sigma)g^{03}\dot{r}_{H}n+2g^{13}\{2\kappa[r-r_{H}]+1\}n-2g^{22}r_{H}^{'2}P_{\theta_{\ast}}\}\frac{\partial S}{\partial r_{\ast}}\\[1mm]
&&\nonumber+[(1-4m\sigma)g^{00}\omega^{2}-2(1-m\sigma)g^{03}\omega n+g^{22}P_{\theta_{\ast}}^{2}+g^{33}n^{2}+m^{2}-O(\sigma^{2})]2\kappa[r-r_{H}]=0.
\end{eqnarray}
Substituting formula~(\ref{eq21}) into~Eq.~(\ref{eq36}), and using $r\rightarrow r_{H}$ simplify the obtained equation, we can get
\begin{eqnarray}
\label{eq37}
&&\nonumber\frac{(1-4m\sigma)a^{2}\sin^{2}\theta\dot{r}_{H}^{2}+\Delta_{\lambda}\Delta_{\theta}A\{2\kappa[r-r_{H}]+1\}^{2}
+\Delta_{\theta}^{2}r_{H}^{'2}
-2(1-m\sigma)\Delta_{\theta}\sqrt{A}(r^{2}+a^{2})\{2\kappa[r-r_{H}]+1\}\dot{r}_{H}}{2\kappa[r-r_{H}]}\left\{\frac{\partial S}{\partial r_{\ast}}\right\}^{2}\\[1mm]
&&-2\{-(1-4m\sigma)a^{2}\sin^{2}\theta\dot{r}_{H}+(1-m\sigma)\Delta_{\theta}\sqrt{A}(r^{2}+a^{2})\{2\kappa[r-r_{H}]+1\}\}
\omega\frac{\partial S}{\partial r_{\ast}}\\[1mm]
&&\nonumber+2\{-a(1-m\sigma)\dot{r}_{H}n+a\Delta_{\theta}\sqrt{A}\{2\kappa[r-r_{H}]+1\}n-\Delta_{\theta}^{2}Ar_{H}^{'2}P_{\theta_{\ast}}\}
\frac{\partial S}{\partial r_{\ast}}=0.
\end{eqnarray}
Let
\begin{eqnarray}
\label{eq38}
\nonumber A_{0}&=&\frac{(1-4m\sigma)a^{2}\sin^{2}\theta\dot{r}_{H}^{2}+\Delta_{\lambda}\Delta_{\theta}A\{2\kappa[r-r_{H}]+1\}^{2}
+\Delta_{\theta}^{2}r_{H}^{'2}
-2(1-m\sigma)\Delta_{\theta}\sqrt{A}(r^{2}+a^{2})\{2\kappa[r-r_{H}]+1\}\dot{r}_{H}}{2\kappa[r-r_{H}]}\\[1mm]
B_{0}&=&\{-(1-4m\sigma)a^{2}\sin^{2}\theta\dot{r}_{H}+(1-m\sigma)\Delta_{\theta}\sqrt{A}(r^{2}+a^{2})\{2\kappa[r-r_{H}]+1\}\}\\[1mm]
\nonumber C_{0}&=&\{-a(1-m\sigma)\dot{r}_{H}n+a\Delta_{\theta}\sqrt{A}\{2\kappa[r-r_{H}]+1\}n-\Delta_{\theta}^{2}Ar_{H}^{'2}P_{\theta_{\ast}}\}.
\end{eqnarray}
Dividing both sides of~Eq.~(\ref{eq37}) by B, then~Eq.~(\ref{eq38}) becomes
\begin{eqnarray}
\label{eq39}
&&\frac{A_{0}}{B_{0}}\left\{\frac{\partial S}{\partial r_{\ast}}\right\}^{2}-2\omega\frac{\partial S}{\partial r_{\ast}}
+2\frac{\partial S}{\partial r_{\ast}}\frac{\partial S}{\partial r_{\ast}}=0.
\end{eqnarray}
When $r\rightarrow r_{H}$ £¬the limit of the coefficient of $\frac{\partial S}{\partial r_{\ast}}^{2}$ at the event horizon should be equal to one, that is
\begin{eqnarray}
\label{eq40}
\lim _{r\rightarrow r_{H}} \frac{A_{0}}{B_{0}}=\frac{(1-4m\sigma)a^{2}\sin^{2}\theta\dot{r}_{H}^{2}+\Delta_{\lambda}\Delta_{\theta}A\{2\kappa[r-r_{H}]+1\}^{2}
+\Delta_{\theta}^{2}r_{H}^{'2}
-2(1-m\sigma)\Delta_{\theta}\sqrt{A}(r^{2}+a^{2})\{2\kappa[r-r_{H}]+1\}\dot{r}_{H}}{2\kappa[r-r_{H}]\{-(1-4m\sigma)a^{2}\sin^{2}
\theta\dot{r}_{H}+(1-m\sigma)\Delta_{\theta}\sqrt{A}(r^{2}+a^{2})\{2\kappa[r-r_{H}]+1\}\}}=1.
\end{eqnarray}
In formula~(\ref{eq40}), the limit of the denominator is zero when $r\rightarrow r_{H}$ , so the limit of the numerator should also be zero when $r\rightarrow r_{H}$ . Using L'hopital's rule, we can work out $\kappa$ as
\begin{eqnarray}
\label{eq41}
\kappa=\frac{(r_{H}-M-\frac{2}{3}\Lambda r_{H}^{3}-\frac{1}{3}\Lambda a^{2}r_{H})\Delta_{\theta}A-2(1-m\sigma)\sqrt{A}\Delta_{\theta}r_{H}\dot{r}_{H}}
{(r_{H}^{2}+a^{2})\sqrt{A}\Delta_{\theta}
+2(r_{H}^{2}+a^{2})\sqrt{A}\Delta_{\theta}\dot{r}_{H}-a^{2}\sin^{2}\theta\dot{r}_{H}-2\Delta_{\theta}A\Delta_{\lambda}}
[1+\sigma \tilde{m}-(\sigma \tilde{m})^{2}+\cdots].
\end{eqnarray}
where
\begin{eqnarray}
\label{eq41-2}
\tilde{m}=\frac{m[(r_{H}^{2}+a^{2})\sqrt{A}\Delta_{\theta}+2(r_{H}^{2}+a^{2})\sqrt{A}\Delta_{\theta}\dot{r}_{H}-4a^{2}\sin^{2}\theta \dot{r}_{H}]}
{(r_{H}^{2}+a^{2})\sqrt{A}\Delta_{\theta}+2(r_{H}^{2}+a^{2})\sqrt{A}\Delta_{\theta}\dot{r}_{H}-a^{2}\sin^{2}\theta \dot{r}_{H}-2\Delta_{\theta}A\Delta_{\lambda}}.
\end{eqnarray}
When in the Schwarzschild space-time, it can be proved that $\tilde{m}=m$ . $\kappa$ is the precisely corrected surface gravity at the event horizon. Making $\lim_{r\rightarrow r_{H}}\frac{C_{0}}{B_{0}}=\omega_{0}$ , we can get
\begin{eqnarray}
\label{eq42}
\omega_{0}=\frac{\Delta_{\theta}\sqrt{A}an+\Delta_{\theta}r_{H}^{'}P_{\theta^{\ast}}-(1-m\sigma)a\dot{r}_{H}n}
{\Delta_{\theta}\sqrt{A}(r_{H}^{2}+a^{2})-a^{2}\sin^{2}\theta\dot{r}_{H}}[1+\sigma m'-(\sigma m')^{2}+\cdots],
\end{eqnarray}
where
\begin{multicols}{2}
\begin{eqnarray}
\label{eq42-2}
m'=\frac{m[\Delta_{\theta}\sqrt{A}(r_{H}^{2}+a^{2})-4a^{2}\sin^{2}\theta\dot{r}_{H}]}
{\Delta_{\theta}\sqrt{A}(r_{H}^{2}+a^{2})-a^{2}\sin^{2}\theta\dot{r}_{H}}.
\end{eqnarray}
$\omega_{0}$ is the exact corrected chemical potential. In combination with~Eq.~(\ref{eq40}) and~(\ref{eq42}),~Eq.~(\ref{eq39}) can be written as follows
\begin{eqnarray}
\label{eq43}
\left\{\frac{\partial S}{\partial r_{\ast}}\right\}^{2}-(\omega-\omega_{0})\left\{\frac{\partial S}{\partial r_{\ast}}\right\}=0.
\end{eqnarray}
Substituting $\frac{\partial s}{\partial r}=\frac{\partial s}{\partial r_{\ast}}\frac{2\kappa(r-r_{H})+1}{\kappa(r-r_{H})}$ in~Eq.~(\ref{eq31}) into~Eq.~(\ref{eq43}), and we can get
\begin{eqnarray}
\label{eq44}
\left\{\frac{\partial S}{\partial r}\right\}_{\pm}=[(\omega-\omega_{0})\pm(\omega-\omega_{0})]\frac{\kappa(r-r_{H})+1}{\kappa(r-r_{H})}.
\end{eqnarray}
When $r\rightarrow r_{H}$ , the residue theorem can be applied to obtain
\begin{eqnarray}
\label{eq45}
\nonumber S_{\pm}&=&\lim_{r\rightarrow r_{H}}\vint[(\omega-\omega_{0})\pm(\omega-\omega_{0})]\frac{\kappa(r-r_{H})+1}{\kappa(r-r_{H})}dr\\[1mm]
&=&\pi i\frac{(\omega-\omega_{0})\pm(\omega-\omega_{0})}{2\kappa}.
\end{eqnarray}
According to tunneling theory, we can obtain the tunneling probability of arbitrary spin fermion in the space-time of Kerr-de Sitter black hole
\begin{eqnarray}
\label{eq46}
\nonumber \Gamma &=&\exp[-2(Im S_{+}-Im S_{-})]\\[1mm]
&=&\exp\left\{-\frac{2\pi(\omega-\omega_{0})}{\kappa}\right\}\\[1mm]
\nonumber &=&\exp\left\{-\frac{\omega-\omega_{0}}{T_{H}}\right\}.
\end{eqnarray}
where
\end{multicols}
\ruleup
\begin{equation}
\label{eq47}
T_{H}=\frac{1}{2\pi}\frac{(r_{H}-M-\frac{2}{3}\Lambda r_{H}^{3}-\frac{1}{3}\Lambda a^{2}r_{H})\Delta_{\theta}A-2(1-m\sigma)\sqrt{A}\Delta_{\theta}r_{H}\dot{r}_{H}}
{(r_{H}^{2}+a^{2})\sqrt{A}\Delta_{\theta}
+2(r_{H}^{2}+a^{2})\sqrt{A}\Delta_{\theta}\dot{r}_{H}-a^{2}\sin^{2}\theta\dot{r_{H}}-2\Delta_{\theta}A\Delta_{\lambda}}
[1+\sigma \tilde{m}-(\sigma \tilde{m})^{2}+\cdots].
\end{equation}
\begin{multicols}{2}
\vspace{0.5cm}
Formula~(\ref{eq47}), $T_{H}$ is the precisely corrected Hawking temperature at the event horizon of the black hole. This is a new form of Hawking temperature expression for the Kerr-de Sitter black hole.

From the demonstration in this section, we can see that through the modified R-S-H-J equation based on the correction of the Lorentz dispersion relation, we obtain a series of precisely corrected physical quantities of Kerr-de Sitter black hole, including surface gravity, chemical potential, tunneling probability and Hawking temperature. The correction is indicated by the parameter. From formulas~(\ref{eq41}),~(\ref{eq42}),~(\ref{eq46}) and~(\ref{eq47}), it can be seen that for a non-stationary Kerr-de Sitter black hole, the surface gravity, chemical potential, tunneling probability and Hawking temperature change with time as the event horizon surface changes. As you can see, the chemical potential also varies with angle $\theta$.\\

\section{Discussion and conclusion}
In this paper, based on the modified Lorentz dispersion relation on the quantum scale and the condition of  $\alpha=2$ selected, we derived the modified R-S-H-J equation by properly selecting the transformation matrix and using the semi-classical method. By using the derived R-S-H-J equation, we solved the dynamic Kerr-de Sitter black hole through the tortoise coordinate transformation, and obtained the accurately corrected surface gravity, chemical potential, tunneling probability and Hawking temperature. Although the correction term $\sigma$ is a small quantity, it is still worth further studying.

Another important physical quantity in the thermodynamics of black hole is black hole entropy. According to the First Law of Thermodynamics, the entropy $S^{s}$of a black hole can be expressed as
\begin{eqnarray}
\label{eq49}
dM=TdS^{s}+VdJ+UdQ.
\end{eqnarray}
For the Kerr-de Sitter black hole,
\begin{eqnarray}
\label{eq50}
dS^{s}=\frac{dM-VdJ}{T}.
\end{eqnarray}
The exactly corrected entropy at the event horizon $r=r_{H}$ of the black hole can be expressed as
\begin{eqnarray}
\label{eq51}
S^{s}_{r_{H}}&=&\vint_{r\rightarrow r_{H}}\frac{dM-VdJ}{T_{H}}\\[1mm]
\nonumber&=&\vint_{r\rightarrow r_{H}}\frac{dM-VdJ}{T_{0}}(1-\sigma \tilde{m})[1-\sigma \tilde{\tilde{m}}+(\sigma \tilde{\tilde{m}})^{2}-\cdots],
\end{eqnarray}
where
\begin{eqnarray}
\label{eq51-2}
\tilde{\tilde{m}}=\frac{2m\sqrt{A}\Delta_{\theta}r_{H}\dot{r}_{H}}{(r_{H}-M-\frac{2}{3}\Lambda r_{H}^{3}-\frac{1}{3}\Lambda a^{2}r_{H})\Delta_{\theta}A-2\sqrt{A}\Delta_{\theta}r_{H}\dot{r}_{H}}.
\end{eqnarray}
In~Eq.~(\ref{eq50}) and~Eq.~(\ref{eq51}), the relational formula $dS_{0}^{s}=\frac{dM-VdJ}{T_{0}}$ represents the black hole entropy before correction, and   represents the Hawking temperature before correction, which is
\begin{eqnarray}
\label{eq52}
T_{0}=\frac{(r_{H}-M-\frac{2}{3}\Lambda r_{H}^{3}-\frac{1}{3}\Lambda a^{2}r_{H})\Delta_{\theta}A-2\sqrt{A}\Delta_{\theta}r_{H}\dot{r}_{H}}
{2\pi[\Delta_{\theta}\sqrt{A}(r_{H}^{2}+a^{2})+a^{2}\sin^{2}\theta \dot{r}_{H}]}.
\end{eqnarray}

Formulas~(\ref{eq41}),~(\ref{eq42}),~(\ref{eq46}) and~(\ref{eq47}) are all new expressions with precise correction. Lorentz dispersion relation are worthy theories to be studied in the field of high energy, which are relations to be considered in the study on the theory of strong gravitational field and gravitational waves. It is worth pointing out that in Lorentz dispersion relations, we take $\alpha=2 $ to study, and the case of  taking other values is also worth further studying, which will be studied in our future work.

\end{multicols}

\vspace{-1mm}
\centerline{\rule{80mm}{0.1pt}}
\vspace{2mm}

\begin{multicols}{2}

\end{multicols}

\clearpage
\end{CJK*}
\end{document}